# Genomic reproducibility in the bioinformatics era

**Pelin Icer Baykal[1,2], Pawel P. Łabaj[3,4], Florian Markowetz[5,6], Lynn M. Schriml[7], Daniel J. Stekhoven[2,8], Serghei Mangul[9,10,*], and Niko Beerenwinkel[1,2,*]**

[1]Department of Biosystems Science and Engineering, ETH Zurich, Basel, 4058, Switzerland
[2]SIB Swiss Institute of Bioinformatics, Basel, 4058, Switzerland
[3]Małopolska Centre of Biotechnology, Jagiellonian University, Gronostajowa 7A, 30-387, Krakow, Poland
[4]Department of Biotechnology, Boku University Vienna, Muthgasse 18, 1190 Vienna, Austria
[5]University of Cambridge, Cambridge, United Kingdom
[6]Cancer Research UK Cambridge Institute, Cambridge, United Kingdom
[7]University of Maryland School of Medicine, Institute for Genome Sciences, 670 W. Baltimore St., HSFIII, Baltimore, MD 21201, USA
[8]NEXUS Personalized Health Technologies, ETH Zurich, 8952 Zurich, Switzerland
[9]Titus Family Department of Clinical Pharmacy, USC Alfred E. Mann School of Pharmacy and Pharmaceutical Sciences, University of Southern California, 1540 Alcazar Street, Los Angeles, CA 90033, USA
[10]Department of Quantitative and Computational Biology, University of Southern California Dornsife College of Letters, Arts, and Sciences, Los Angeles, CA 90089, USA
*Corresponding authors.

## Abstract

In biomedical research, validation of a new scientific discovery is tied to the reproducibility of its experimental results. However, in genomics, the definition and implementation of reproducibility still remain imprecise. Here, we argue that genomic reproducibility, defined as the ability of bioinformatics tools to maintain consistent genomics results across technical replicates, is key to generating scientific knowledge and enabling medical applications. We first discuss different concepts of reproducibility and then focus on reproducibility in the context of genomics, aiming to establish clear definitions of relevant terms. We then focus on the role of bioinformatics tools and their impact on genomic reproducibility and assess methods of evaluating bioinformatics tools in terms of genomic reproducibility. Lastly, we suggest best practices for enhancing genomic reproducibility, with an emphasis on assessing the performance of bioinformatics tools through rigorous testing across multiple technical replicates.

## Introduction

Recent advancements in genomics, specifically in sequencing technologies and bioinformatics, have paved the road to precision medicine [1]. The ability to analyze an individual's genetic information has opened up new possibilities for tailored treatments and improved patient outcomes. However, to ensure the credibility and progress of genomic medicine, the reproducibility of results across laboratories has emerged as a crucial limiting factor.

The multifaceted nature of reproducibility in genomics research is due to the various steps involved in the production of genomic data and results. These steps encompass experimental procedures, which include sample preparation and sequencing, and data analysis tasks, such as aligning reads to a reference genome, calling variants, and analyzing gene expression levels. The experimental variability occurring during the production of genomic data poses a considerable challenge for bioinformatics tools, as they are supposed to generate consistent results under such variation.

Consistency in genomic results is a fundamental requirement for bioinformatics tools when applied to identical genomic data. This aspect is commonly referred to as methods reproducibility in experimental studies [2]. Methods reproducibility, as defined by Goodman et al., pertains to the ability of precisely executing, to the highest degree possible, the experimental and computational procedures, using the same data and tools, in order to yield identical results [2]. In the context of genomics, methods reproducibility refers to obtaining the same results across multiple runs of the bioinformatics tools using the same parameters and genomic data (Fig. 1). Ideally, bioinformatics tools should also provide consistent results when analyzing genomic data obtained from different sequencing runs, including in different laboratories, but using the same protocols. A single, universally recognized term that describes the impact of bioinformatics tools on genomic results across such technical replicates is currently lacking. Pan et al., discuss reproducibility in the context of specific bioinformatics tasks. For instance, the reproducibility impact of read alignment tools is referred to as "aligner reproducibility," while the reproducibility of structural variant callers is termed "caller reproducibility" [3]. The authors assess the consistency of these bioinformatics tasks across multiple tools and datasets. The closest definitions for this assessment were introduced by Goodman et al. [2] as results reproducibility and by Gundersen [4] as outcome reproducibility. Results reproducibility is the ability to obtain the same results when independent studies on different datasets are conducted with procedures closely resembling the original study [2]. However, the concept of results reproducibility was defined to target the reproduction of an experiment including a handful of statistical tests, rather than the analysis of high-dimensional and heterogeneous multi-omics data produced regularly by large collaborative genomics initiatives today. Therefore, we propose the term genomic reproducibility for the ability to obtain consistent outcomes from bioinformatics tools using genomic data obtained from different library preparations and sequencing runs, but for fixed experimental protocols (Fig. 1).

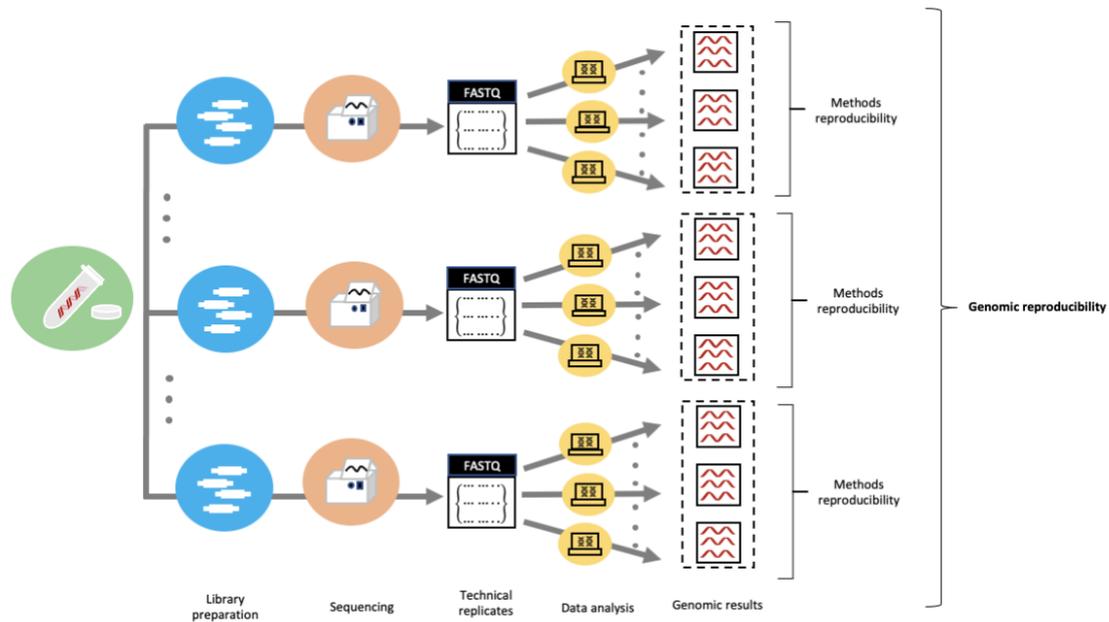

**Figure 1:** Schematic representation of three key concepts: technical replicates, methods reproducibility, and genomic reproducibility. The same sample is processed (library preparation) and sequenced multiple times, possibly in different laboratories, but using the same experimental protocols and sequencing platform. The output of these sequencing runs are technical replicates represented as FASTQ files. Data analysis is performed for each technical replicate multiple times to assess consistency of genomic results, which refers to methods reproducibility. Genomic reproducibility, on the other hand, evaluates the consistency of genomic results across technical replicates.

## Reproducibility in genomics

Genomic reproducibility can be jeopardized during two critical stages: the pre-sequencing and sequencing steps where technical variability may arise, and the computational analysis and interpretation of the genomic data, often due to the impact of stochastic algorithms. In the context of DNA sequencing, technical variability can arise from the use of diverse sequencing platforms [5] and from differences between individual flow cells [3, 6, 7]. Even if the sequencing protocol is kept identical across multiple runs, experimental variation is still expected as a result of the random sampling variance of the sequencing process and variations in library preparation [8-10]. In light of this, the objective of bioinformatics tools should be to accommodate and tolerate such experimental variation, aiming to generate consistent results across different sequencing runs and library preparations, which means achieving genomic reproducibility.

In practice, controlling conditions of sequencing experiments is challenging and high levels of experimental variations may compromise the ability of bioinformatics tools to maintain consistent results across technical replicates. In order to evaluate the performance of bioinformatics tools in terms of genomic reproducibility, one can consider technical replicates that capture specifically the variations among sequencing runs and library preparation techniques. This approach intentionally disregards other potential factors that could confound the results, such as sequencing protocols and platforms, allowing technical replicates acquired under the exact same sequencing protocols to be utilized for evaluating the impact of bioinformatics tools. However, generating technical replicates can escalate both the financial burden and logistical complexity of genomic experiments, and in certain cases, it may be impractical or ethically prohibitive to obtain them, particularly in clinical settings.

## Bioinformatics tools can remove but also introduce unwanted variation

Bioinformatics tools play a crucial role in analyzing, interpreting, and eliminating undesired variation in genomic data. Variations in genomic data can arise due to multiple sources such as experimental noise, sequencing errors, or biological artifacts. To ensure that systematic errors do not confound the results, bioinformatics tools employ techniques like normalization to remove batch effects or technical biases from genomic data [11]. However, despite their indispensable nature, bioinformatics tools are not perfect, and in certain cases, they may actually introduce unwanted variation, which can significantly impact the accuracy of genomic data analyses. The introduction of unwanted variation may arise from various sources, including algorithmic biases, data preprocessing techniques, or misuse of tool parameters. Below we focus on read alignment and structural variant tools to discuss this phenomenon and its impact on genomic reproducibility in more detail.

An example of algorithmic bias is reference bias, which occurs when alignment algorithms favor aligning reads to a specific reference genome sequence. Lunter and Goodson demonstrated in their analysis that some alignment tools (BWA [12] and Stampy [13]) consistently show a bias in mapping reads containing reference alleles of a known heterozygous indel in the genome [13]. When it comes to data processing techniques, setting a low threshold for quality filtering can lead to undesired variations in read alignment by including low-quality reads with sequencing errors. Since structural variant callers rely on the input from read alignment tools, these variations also impact the detection of structural variants [14]. These examples illustrate the potential impact of bioinformatics tools on the variability of genomic results. Gaining an understanding of the types and amount of the variability introduced by bioinformatics tools is an essential first step to achieving genomic reproducibility.

## Genomic reproducibility of read alignment tools and variant callers

Stochastic data analysis algorithms are an obvious source of variation of genomic results [15]. In general, they yield different results, even for the exact same input data. For example, one of the challenges of read alignment tools is to capture and report reads mapped to repetitive regions of the reference genome, which are known as multi-mapped reads [16]. There exist different strategies to deal with the uncertainty of multi-mapped reads: some tools do not take these reads into account at all (e.g., SNAP [17]), other tools use a deterministic approach and try to identify the best possible position among all the matching positions (e.g., RazerS [18] and mrFAST [19]), and finally some tools adopt a non-deterministic approach and select random a position among all the matching positions (e.g., BWA-MEM [12] and Stampy [13]).

Allowing users to set a seed for a pseudo-random generator can, in the case of multi-mapping, restore reproducibility of stochastic alignment strategies (Table 2). However, some tools employ hashing methods that link read attributes to the read mapping positions. For instance, Bowtie2 [20] uses the name of the reads and the associated mapping positions as pairs for hashing. On the other hand, BWA-MEM [12] uses the position index of reads within a file to pair with mapping positions of the reads. Since the chosen multi-mapping position is bound to the order of its associated read, changing only the order of the reads leads, in general, to different results [21].

According to one study, which benchmarked several variant callers and read alignment tools, reports of structural variants vary across different variant callers and among the same variant callers when different read alignment tools are used [3]. In another study, Alkan et al. reported that the structural variant calling tools DELLY [22], LUMPY [23], and Genome STRiP [24] produced 3.5 to 25.0% of different variant call sets with randomly shuffled data compared to the original data [21]. These studies demonstrate the potential impact of bioinformatics algorithms on the reproducibility of genomic results and emphasize the significance of assessing it with replicates (Table 2).

## Opportunities to assess the impact of bioinformatics tools on genomic reproducibility

Ongoing efforts in genomics include ensuring whole-genome sequencing (WGS) reproducibility, with notable initiatives including the Genome in a Bottle (GIAB) consortium, hosted by the National Institute of Standards and Technology (NIST), and the HapMap project. The complementing efforts were performed within consecutive phases of the US FDA-led MicroArray/Sequencing Quality Control Project (MAQC/SEQC), which is helping improve microarray and next-generation sequencing technologies and foster their proper applications in discovery, development, and review of FDA-regulated products. In the MAQC-IV/SEQC phase, the aim was to assess the technical performance of next-generation sequencing platforms by generating benchmark datasets with reference samples and evaluating advantages and limitations of various bioinformatics strategies in RNA and DNA analyses. The impact of various bioinformatics approaches on the downstream biological interpretations of RNA-seq results was comprehensively examined and the utility of RNA-seq in clinical application and safety evaluation was assessed. In SEQC2, which is the next phase of SEQC, the focus has been placed on targeted DNA- and RNA-seq to develop standard analysis protocols and quality control metrics for fit-for-purpose use of NGS data to enhance regulatory science research and precision medicine. On the other hand, consortiums such as the GIAB, and the HapMap projects provide reference materials that are used to evaluate genomic reproducibility in various studies. In Table 1, DNA and RNA-seq technical replicate datasets from major consortiums and studies are compiled, which can be used to assess genomic reproducibility.

Technical replicates of the Ashkenazi Trio dataset were generated to assess the performance of DNA sequencing platforms [5]. This involved generating triplicates of inter-laboratory and intra-laboratory paired-end and single-end DNA-seq samples using five Illumina and three ThermoFisher Ion Torrent platforms. This dataset can serve as a valuable resource for assessing genomic reproducibility by examining the performance of DNA-seq alignment tools and structural variant tools using both paired-end and single-end triplicate samples. The Chinese Quartet dataset, the HapMap Trio, and a pilot genome NA12878 are datasets with technical replicates that have been generated for structural variant detection studies [3, 25]. Pan et al. used technical replicates from the Chinese Quartet to assess reproducibility across three different labs using different alignment and structural variant callers [3, 25]. These technical replicates were sequenced from three different labs as triplicates representing different runs of sequencing. The same dataset was used to evaluate how sequencing centers, replicates, alignment tools and platforms affect SV calling in NGS [25]. Additionally, The HapMap Trio and the NA12878 datasets were employed in a separate SV calling study to examine reproducibility across various factors, including sequencing platforms, labs, library preparations, alignment tools, and SV calling tools [3]. Technical replicates consist of triplicates of short-reads which can again be used to assess genomic reproducibility and the findings can be compared to the findings available in SV calling studies [3, 25]. Lastly, we mention an RNA-seq dataset provided by the SEQC consortium [26], which has been employed to assess the reproducibility of RNA-seq experiments [11, 27] and also the impact of RNA-seq data analysis tools on gene expression analysis [15]. Four samples were sequenced in 4 technical replicates each, then the whole experiment was replicated in 6 different sites all over the world and another 5th replicate was created by a vendor and sent to labs for sequencing. All RNA-seq technical replicates used in these studies are made publicly available, serving as a valuable resource for assessing genomic reproducibility.

**Table1:** Technical replicates obtained from selected genomics consortia. The "Reference material" column indicates the reference material name used to generate technical replicates. The "Consortium" column specifies the consortium responsible for obtaining patient consent or providing the reference materials. The "Data Type" column indicates the specific type of data associated with the reference material. In the "Technical replicates properties" column, details regarding technical replicates are presented. The "Accession" column provides information on the platform and identification numbers used to access the technical replicates. Finally, the "Study" column references the original study where these technical replicates were generated.

| Reference material | Consortium | Data type | Technical replicates properties | Accession | Link to data | Study |
|---|---|---|---|---|---|---|
| Ashkenazi Jewish Trio NIST IDs: HG002 (son) HG003 (father) HG004 (mother) | Personal Genome Project (PGP) [33] | WGS | Inter- and intra-laboratory Multiple platforms | Available in BioProject PRJNA646948 With SRA: SRR12898279–SRR12898354 | https://www.ncbi.nlm.nih.gov/bioproject/?term=PRJNA646948 | [5] |
| Chinese Quartet IDs: LCL5 & LCL 6 (monozygotic twins) LCL7 & LCL8 (parents) | The Quartet Project for Quality Control and Data Integration of Multi-omics Profiling | WGS | Inter- and intra-laboratory Multiple platforms | Available in NODE OEP001896 | biosino.org/node/project/detail/OEP001896 | [3, 25] |
| | http://chinese-quartet.org | | | | | |
| HapMap Trio IDs: NA10385 NA12248 NA12878 | The International HapMap Project [34] | WGS | Inter- and intra-laboratory Multiple platforms | Available in BioProject PRJNA723125 | https://www.ncbi.nlm.nih.gov/bioproject/PRJNA723125 | [3] |
| NA12878 (HG001) (Pilot genome) | Genome in a Bottle (GIAB) led by the NIST [35] | WGS | Inter- and intra-laboratory Multiple platforms | NA | NA | [3] |
| MAQC-I - microarrays: Universal Human RNA Reference (UHRR) and Human Brain RNA Reference (HBRR) MAQC-III (SEQC) - same samples with RNA-SEQ MAQC-IV/SEQC2 - targeted RNA-Seq but also | MAQC-I MAQC-III MAQC-IV/SEQC2 | Microarrays RNA-Seq Targeted RNA-Seq and Targeted DNA-Seq | Inter- and intra-laboratory Multiple platforms | Available in Gene Expression Omnibus (GEO) MAQC-I: GSE5350 SEQC: GSE47792 (SuperSeries) contains GSE47774 (just RNA-Seq) SEQC2: targeted RNA-Seq [not published yet] Targeted DNA-Seq [ https://www.ncbi. | https://www.ncbi.nlm.nih.gov/geo/query/acc.cgi?acc=GSE47774 | [26] |
| targeted DNA-Seq - UHRR plus normal male cell line (Agilent Human Reference DNA, Male, Agilent part #: 5190-8848) | | | | nlm.nih.gov/bioproject/PRJNA677997] | | |

## Synthetic replicates

In certain conditions, such as when the number of technical replicates is limited for a specific type of genomic data or when reproducibility assessment requires a more controlled environment, synthetic replicates may be employed instead of technical replicates. This approach allows for a more controlled examination of the impact of specific alterations in the data. Synthetic replicates are generated *in silico* to mimic the variations of sequencing output expected from technical replicates. In practice, it is not possible to computationally reproduce all variations among technical replicates, but there exist different techniques to generate synthetic replicates that reflect some of the variations.

One approach to create synthetic replicates is randomly shuffling the order of the reads reported from a sequencer (Fig. 2, top), which reflects the randomness of events in a sequencing experiment, such as DNA hybridization on the flow cell [21]. Another technique is to take the reverse complement of each read (Fig. 2, middle) to assess strand bias [28] when the reference genome is double-stranded. The bias arises due to a pronounced overabundance in one direction of NGS sequencing reads either forward or reverse, compared to the opposite direction [29]. This problem may lead to unwanted variation which can impact genomic reproducibility. Yet another technique is bootstrapping (Fig. 2, bottom) reflecting random sampling variance, which is a widely used type of synthetic replicate employed in many genomics, transcriptomics [30] and metagenomics [31] studies.

Both technical replicates and synthetic replicates have their own advantages and limitations. Technical replicates contribute to a more realistic and reliable assessment by accounting for inherent variability in experimental procedures, such as different sequencing runs, and enabling rigorous statistical analysis. On the other hand, synthetic replicates offer a controlled evaluation of tools since the modifications applied to the data are known, allowing for a precise assessment against a ground truth. Hence, utilizing both types of replicates can be useful in assessing genomic reproducibility.

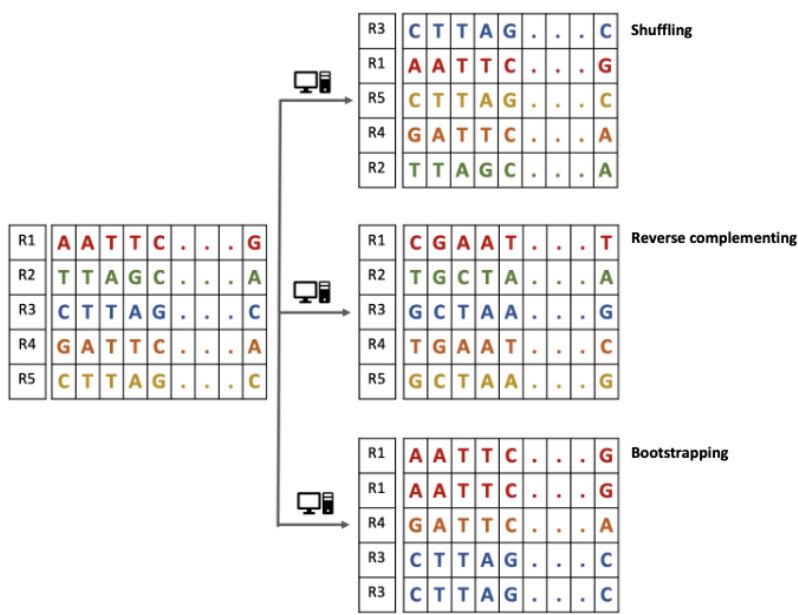

**Figure 2:** Schematic representation of generating synthetic replicates. Based on a given dataset consisting of five reads R1, …, R5 (left) three different types of synthetic replicates (right) are created by either randomly shuffling the order of the five reads (top), or by taking the reverse complement of each read (middle), or by bootstrapping, i.e., resampling of the five reads with replacement (bottom).

## Best practices to improve genomic reproducibility

We have compiled a set of recommended standards and guidelines aimed at promoting genomic reproducibility. These recommendations are based on the assumption that bioinformatics tools already adhere to guidelines concerning technical challenges associated with operating systems, hardware and workflow management systems, which are essential for ensuring methods reproducibility [32].

We suggest the following best practices for the development and application of bioinformatics tools to ensure genomic reproducibility. First, tools should be documented sufficiently including detailed explanations of all parameters and their default settings, and usage examples and guidelines to assist users in selecting appropriate parameter values. Furthermore, tool developers should clarify in the documentation the relationship between parameter selection and reproducibility to facilitate accurate and consistent results.

The second essential requirement involves incorporating functionality that allows users to specify random seeds. By implementing this feature, developers provide users control over the random results generated by non-deterministic algorithms. This control is vital for ensuring that the same set of input data consistently produces the same output, enabling to assess methods reproducibility.

Another recommendation pertains to the performance assessment of the bioinformatics tool. It is essential to conduct controlled experiments using synthetic replicates, technical replicates, or a combination of both. The result obtained from these experiments, along with any observed discrepancies or variations, should be thoroughly reported. This comprehensive reporting enables researchers to evaluate the performance and reliability of the tool accurately.

**Table 2:** Recommended genomic reproducibility standards. The "Standard" column lists the name of the standards aimed at ensuring genomic reproducibility. The "Guideline" column describes the methodologies for attaining the respective standard. The columns "Essential" and "Desirable" columns categorize the levels of significance attached to each individual standard.

| Standard | Guideline | Essential | Desirable |
|---|---|---|---|
| Documentation | • Document all the parameters of the tool, including their names, descriptions, acceptable values, and default settings. | x | |

| | | | x |
|---|---|---|---|
| | • Provide detailed explanations of each parameter and its impact on the analysis or processing. | | |
| | • Include usage examples and guidelines to help users choose appropriate parameter values. | | |
| | • Highlight the relationship between parameter selection and reproducibility. | | |
| Random seeds | • Implement functionality to define random seeds for any random process involved. | x | |
| | • Document how specified random seeds impact reproducibility. | | |
| | • Provide examples for selecting appropriate random seeds to ensure reproducibility. | | |
| Assessment of reproducibility | • Conduct a controlled experiment using synthetic replicates or technical replicates or ideally both. | x | |
| | • Report results obtained from the replicates, including any observed discrepancies or variations. | | |
| Visualization of reproducibility performance | • Generate visual representations, such as plot, heatmaps etc. to examine results obtained from replicates. | | x |
| | • Clearly describe the purpose and interpretation of each visualization. | | |

Finally, bioinformatics tool developers can enhance reproducibility by providing result visualization from replicates. However, effectively handling visualization and communicating results poses challenges due to extensive scale and complexity of the genomic data involved. By employing suitable visualization techniques and dimensionality reduction methods, these challenges can be overcome. Through careful analysis of patterns of discrepancies from the visualizations, researchers can gain valuable insights into the reliability and consistency of the results produced by the tool.

## Conclusion

Reproducibility is critical in all fields of science, engineering, and medicine, to ensure the reliability and integrity of findings and the safeness of their applications. However, there are various challenges and limitations to achieving reproducibility in practice. The field of genomics faces several hurdles to reproducibility due to rapid advancements in sequencing technologies and data generation. Each new technology introduces unique biases and sources of variation, which need to be carefully considered and addressed during data analysis. Additionally, genomic studies often involve complex bioinformatics pipelines, which are susceptible to errors and require rigorous validation.

Bioinformatics tools have made significant contributions to mitigating some of these challenges and enhancing genomic reproducibility. These tools facilitate the standardization and automation of data processing, analysis and visualizations, minimizing human error and increasing the reliability of results. However, bioinformatics tools are not without limitations and can even introduce unwanted variations that compromise genomic reproducibility. Using technical and synthetic replicates present valuable approaches for evaluating essential aspects of bioinformatics algorithms and their influence on genomic reproducibility.

The use of technical replicates offers advantages, as it captures the diversity across different runs of sequencing. In order to correctly assess bioinformatics tools in terms of genomic reproducibility, it is important to acknowledge that despite efforts to control experimental conditions, variations can arise due to factors such as human errors in sample preparations or unknown batch effects. These confounding factors and other experimental parameters such as variations in sequencing platforms can influence genomic results. We recommend the use of technical replicates to capture variations arising from different runs of sequencing and different library preparations. Additionally, it is vital to grasp the degree to which non-deterministic algorithms can influence genomic results and then tailor the assessment of genomic reproducibility accordingly.

Synthetic replicates are a fast and cost-efficient way of generating replicates in genomics. They cannot fully represent real data variation as they capture only some of the differences produced between different runs of sequencing. However, they provide a useful and easily accessible way of early assessing necessary features of bioinformatics algorithms and the way they impact on genomic reproducibility.

Precision medicine heavily relies on accurate and reliable genomic information. However, the reliability of genomic results can only be ensured if they are reproducible by bioinformatics tools. As such, it is essential to consider reproducibility as a key evaluation criteria when assessing the quality of these tools. We recommend that both developers and users of bioinformatics tools follow the guidelines in Table 2 to ensure genomic reproducibility. By implementing these guidelines, we can improve the reliability of analyzing genomic data, and facilitate the successful translation of precision medicine to clinical practice.